\documentclass[aps,prb,twocolumn,superscriptaddress,showpacs]{revtex4}

\usepackage{graphicx}
\usepackage{mathrsfs}
\usepackage{bm}
\usepackage{amsmath}
\usepackage{dcolumn}
\usepackage{epstopdf}
\usepackage{dsfont}
\usepackage{amssymb}

\begin{document}
\title{Topological metallic states in spin-orbit coupled bilayer systems}

\author{Hui Pan}
\email{hpan@buaa.edu.cn}
\affiliation{Department of Physics, Beihang University, Beijing 100191, China}
\affiliation{Engineering Product Development, Singapore University of Technology and Design, Singapore 138682, Singapore}

\author{Xin Li}
\affiliation{Department of Physics, Beihang University, Beijing 100191, China}

\author{Zhenhua Qiao}
\affiliation{Department of Physics, University of Science and Technology of China, Hefei, Anhui 230026, China}
\affiliation{ICQD, Hefei National Laboratory for Physical Sciences at Microscale, University of Science and Technology of China, Hefei, Anhui 230026, China}

\author{Cheng-Cheng Liu}
\affiliation{School of Physics, Beijing Institute of Technology, Beijing 100081, China}

\author{Yugui Yao}
\affiliation{School of Physics, Beijing Institute of Technology, Beijing 100081, China}

\author{Shengyuan A. Yang}
\email{shengyuan_yang@sutd.edu.sg}
\affiliation{Engineering Product Development, Singapore University of Technology and Design, Singapore 138682, Singapore}

\begin{abstract}
  We investigate the influence of different spin-orbit couplings on topological phase transitions in the bilayer Kane-Mele model. We find that the competition between intrinsic spin-orbit coupling and Rashba spin-orbit coupling can lead to two dimensional topological metallic states with nontrivial topology. Such phases, although having a metallic bulk, still possess edge states with well-defined topological invariants. Specifically, we show that with preserved time reversal symmetry, the system can exhibit a $\mathbb{Z}_2$-metallic phase with spin helical edge states and a nontrivial $\mathbb{Z}_2$ invariant. When time reversal symmetry is broken, a Chern metallic phase could appear with chiral edge states and a nontrivial Chern invariant.
\end{abstract}


\maketitle

\section{Introduction}

The study of effects of spin-orbit coupling (SOC) has been a central theme in condensed matter physics in the past decade. SOC is an essential ingredient in spintronics, which aims to utilize the electron's spin degree of freedom instead of charge for information processing, with the advantages of smaller size, fast speed and low dissipation. A perhaps more surprising discovery is that SOC can give rise to new quantum states of matter, e.g., topological insulators that are bulk insulators characterized by nontrivial topological invariants and gapless surface (or edge) states. In two dimension (2D), a topological insulator with time reversal symmetry is also known as the quantum spin Hall (QSH)insulator.\cite{Kane1,ZhangSC1} It has an insulating bulk and gapless spin helical edge states protected by a topological $\mathbb{Z}_2$ invariant. When time reversal symmetry is broken, the system may realize a quantum anomalous Hall (QAH)\cite{Onoda,YuR,Nagaosa,QiaoZH1,QiaoZH2} insulator with chiral edge states protected by so-called TKNN or Chern invariant.\cite{Laughlin,Thouless} Their intriguing properties have been a subject of intensive investigations in recent years. Experimentally QSH effect has been demonstrated in HgTe/CdTe quantum wells\cite{ZhangSC2,konig} and inverted InAs/GaSb quantum wells,\cite{liu2008,knez1,knez2} and QAH effect has been demonstrated in Cr-doped (Bi,Sb)$_2$Te$_3$ thin films recently.\cite{ChangCZ} These discoveries further stimulate significant interests in searching for new topological states of matter.

In solid state systems, SOC comes into the Hamiltonian in different forms due to different physical origins. For example, in III-V semiconductor quantum wells, there is Dresselhaus SOC from bulk crystalline inversion symmetry breaking, and there also exists Rashba SOC from structural inversion symmetry breaking along the growth direction. The ability to tune the strength of each individual SOC and control their competition is at the heart of spintronics applications. In the seminal work by Kane and Mele that proposed the 2D $\mathbb{Z}_2$ topological insulator, two types of SOC were considered: the intrinsic SOC and the Rashba SOC.\cite{Kane2} It has been shown that the intrinsic SOC favors the QSH state while the Rashba SOC tends to destroy it. Hence their competition determines the topological phase transitions and the phase boundaries between the topologically trivial and nontrivial insulating states.

The Kane-Mele model was first proposed for graphene. Later studies showed that the magnitude of intrinsic SOC in graphene is too small for the QSH effect to be detected experimentally.\cite{YaoYG11,Brataas,MacDonald3} Recently, a lot of new 2D materials have been discovered with enhanced intrinsic type SOC, such as silicene and germanene\cite{Lalmi,Vogt,Fleurence,ChenL,Tsai,LiuCC1} which are group IV counterparts of graphene and 2D honeycomb Bi/Sb halides and hydrides. The strength of intrinsic SOC can be as large as 0.6eV as in some Bi halides. This provides promising systems to realize the Kane-Mele model and the QSH effect at room temperature.

In the study of 2D physics, bilayer systems have attracted a lot of interest. Compared with single layer system, the extra layer degree of freedom can lead to many new physical phenomena and is usually easier to control in practice.\cite{McCann,MacDonald1,MacDonald2,Morpurgo} Furthermore, in terms of topological properties, bilayer systems can differ qualitatively from single layer systems. For example, when two single layers of QSH insulator are combined to form a bilayer system, its $\mathbb{Z}_2$ invariant vanishes and the system becomes topologically trivial. As another example, it has been found that the Rashba SOC which tends to destroy the QSH phase in single layer turns out to favor QSH phase in bilayer graphene.\cite{QiaoZH3} All these indicate that the physics of bilayer is quite different from the single layer case.

In this work, motivated by the above mentioned progress, we study the effects of different SOCs in a bilayer Kane-Mele model, focusing on its topological properties. We find that the bilayer system exhibits a very rich phase diagram. Quite interestingly, we discover novel 2D topological metallic phases in this system due to the competition between the intrinsic SOC and the Rashba SOC: the 2D $\mathbb{Z}_2$-metallic phase and the Chern metallic phase, depending on whether time reversal symmetry is broken. In such phases, the bulk band gap is closed indirectly, hence a topological invariant can still be well defined, and the state is adiabatically connected to a topologically nontrivial insulator. Like topological insulating phases, the hallmark of the topological metallic phase is the presence of edge states determined by the bulk topology. For $\mathbb{Z}_2$-metallic phase, they are the spin helical edge states, while for Chern metallic phase, they are the chiral edge states. These findings not only extend our understanding of topological states of matter but also may find useful applications based on topological materials.

Our paper is organized as follows. In Section II, we describe the bilayer Kane-Mele model that we study. In Section III, we investigate the case with preserved time reversal symmetry and show that a metallic phase with well-defined $\mathbb{Z}_2$ invariant and spin helical edge states can be realized. Section IV is for the case with time reversal symmetry breaking and we show that there exists a metallic phase with well-defined Chern invariant and chiral edge states. Finally a summary of our work is given in Section V.

\section{Physical model}
\begin{figure}
  \includegraphics[width=7cm]{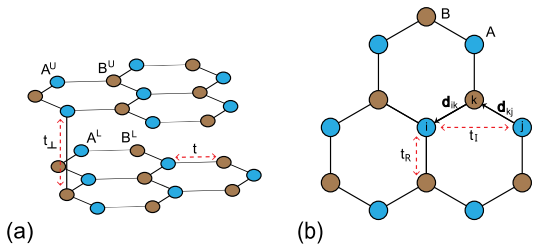}
  \caption{\label{FIG:Structure} (color online) (a) Structure of the bilayer model with the the intralayer (interlayer) hopping parameter $t$ ($t_{\perp}$). The A (B) sublattices are indicated by blue (brown) spheres. (b) Top view of a single layer. $d_{ik}$ and $d_{kj}$ are the nearest-neighbor vectors. $t_\text{I}$ and $t_\text{R}$ are respectively intrinsic SOC and Rahsba SOC strengths. }
\end{figure}

The original Kane-Mele model is written for graphene which has a honeycomb lattice with one orbital per site and two sites per unit cell. For our bilayer model, we take two AB-stacked honeycomb lattices, each described by a Kane-Mele model, as shown in Fig.~\ref{FIG:Structure}. The total Hamiltonian of the system can be written as
\begin{equation}\label{EQ:BilayerH}
\begin{split}
\mathcal{H}&= H^\text{U}+H^\text{L}+t_{\perp}\sum_{i\in \text{(U,A)},j\in \text{(L,B)},\alpha}\left(c^\dagger_{i\alpha}c_{j\alpha}+\text{h.c.}\right) \\
        &+ U\left(\sum_{i\in \text U,\alpha}c^\dagger_{i\alpha}c_{i\alpha}-\sum_{i\in \text L,\alpha}c^\dagger_{i\alpha}c_{i\alpha}\right),
\end{split}
\end{equation}
where $H^\text{U}$ and $H^\text{L}$ are the Hamiltonians for the upper layer and lower layer respectively. The third term is the interlayer coupling. Here due to the stacking geometry, we only consider the hopping between the A site of the upper layer and the nearest B site of the lower layer. $t_\perp$ is the interlayer hopping amplitude. $c$ ($c^\dagger$) is the annihilation (creation) operator for an electron, and h.c. denotes Hermitian conjugate. Subscripts $i$, $j$ label the lattice sites and $\alpha$ is the spin index. The last term represents an interlayer bias potential with strength $U$.
Each single-layer Hamiltonian $H^{\text{U(L)}}$ is a Kane-Mele model containing the following terms
\begin{equation}\label{EQ:SingleH}
H^{\text{U(L)}}=H_\text{hop}+H_\text{ISO}+H_\text{RSO}+H_\text{m},
\end{equation}
where
\begin{eqnarray*}
H_\text{hop} &=& -t\sum_{\langle ij \rangle, \alpha}c^\dagger_{i\alpha}c_{j\alpha}, \\
H_\text{ISO} &=& i t_\text{I}\sum_{\langle\langle ij \rangle\rangle, \alpha\beta}
 \nu_{ij}c^\dagger_{i\alpha}{\sigma}^{z}_{\alpha\beta}c_{j\beta}, \\
H_\text{RSO} &=& i t_\text{R}\sum_{\langle ij \rangle, \alpha\beta}c^\dagger_{i\alpha}
 (\bm{\sigma} \times \hat{\bm{d}}_{ij})^{z}_{\alpha\beta} c_{j\beta}, \\
H_\text{m} &=& M\sum_{i,\alpha\beta}c^\dagger_{i\alpha}{\sigma}^{z}_{\alpha\beta}c_{i\beta}.
\end{eqnarray*}
The first term $H_\text{hop}$ represents the nearest neighbor hopping term with hopping energy $t$ which is used as energy unit in the following. The second term $H_\text{ISO}$ is the intrinsic SOC involving the next-nearest neighbor hopping with amplitude $t_\text{I}$, $\nu_{ij}={\bm{d}_{kj} \times \bm{d}_{ik}/{|\bm{d}_{kj} \times \bm{d}_{ik}|}}=\pm 1$, where $\bm{d}_{lm}$ is the vector along the bond from site $m$ to a nearest site $l$. $\bm \sigma=(\sigma_x,\sigma_y,\sigma_z)$ is the vector of spin Pauli matrices. The summation over $\langle...\rangle$ ($\langle\langle...\rangle\rangle$) runs over all the nearest (next-nearest) neighbor sites.
The third term $H_\text{RSO}$ is the Rahsba SOC with strength $t_\text{R}$, and $\hat{\bm{d}}_{lm}=\bm{d}_{lm}/|\bm{d}_{lm}|$ is a unit vector. The last term $H_\text{m}$ represents a spin-splitting from a Zeeman-like coupling with strength $M$, which can be induced e.g. by magnetic proximity effect.

The intrinsic SOC is from the crystalline structure of the honeycomb lattice. For graphene with a flat planar structure, its value is small because it is a second order process. However, for a low-buckled structure in which A and B sites have a relative shift in the out of plane direction, the intrinsic SOC can be greatly enhanced, as in silicene and germanene. The Rashba SOC here is due to inversion symmetry breaking along the out of plane direction, which may be induced by a substrate or an external electric field along $z$-direction.

When $U=M=0$, the system has both inversion symmetry and time reversal symmetry, the system is a zero gap semiconductor. A finite $U$ breaks inversion symmetry, while a finite $M$ breaks time reversal symmetry, leading to symmetry-breaking ground states. This generally opens a bulk gap. In the following, we shall analyze these symmetry-breaking states in detail.

\section{$\mathbb{Z}_2$-metallic phase}
\begin{figure}
  \includegraphics[width=8cm]{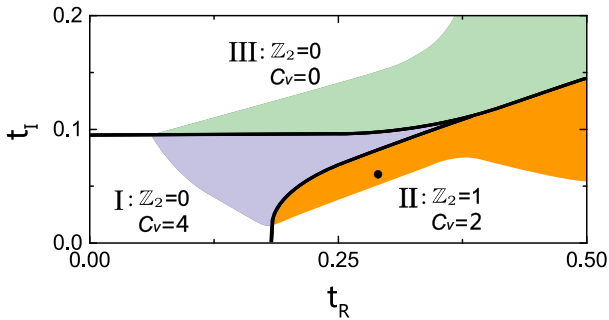}
  \caption{\label{FIG:PhaseISB} (color online) Phase diagram of the bilayer Kane-Mele model as a function of $t_{R}$ and $t_{I}$. The solid lines separate three phases with different topological invariants $\mathbb{Z}_2$ and $\mathcal{C}_v$. The colored regions are the metallic states where the band gap closes indirectly. The parameters used here are $t=1$, $U=0.5$, $M=0$, and $t_{\perp}=0.2$.}
\end{figure}

We first consider the case with preserved time reversal symmetry by setting $M=0$.
In this case, the Chern invariant must vanish. However, the system can still have a nontrivial $\mathbb{Z}_2$ invariant. If the system has preserved spin component, e.g. with only intrinsic SOC but vanishing Rashba SOC, the two spin species have opposite Chern numbers. Then the $\mathbb{Z}_2$ invariant just equals (half) the difference between the Chern numbers of the two spin species $\sigma_z=\pm 1$ (modular 2).
For the general case with no conserved spin components, the $\mathbb{Z}_2$ invariant can be computed using the following formula\cite{Kane2,Hatsugai-Moore}
\begin{equation}\label{EQ:Z2}
\mathbb{Z}_2=\frac{1}{2\pi}\left[\oint_{\partial \text{HBZ}}d\bm{k}\cdot \bm{A}(\bm{k})
 -\int_{\text{HBZ}}d^2 k\,\mathit \Omega_z(\bm{k})\right] \,\, \text{mod}\,2,
\end{equation}
where $\bm A(\bm{k})=i\sum_n \langle u_n(\bm{k})|\nabla_k u_n(\bm{k}) \rangle$ is the Berry connection summed over all the occupied bands, $|u_n(\bm{k})\rangle$ is the periodic part of the Bloch state for band $n$. $\mathit \Omega_z(\bm{k})=(\nabla_k \times \bm{A})_z$ is the $z$-component of the Berry curvature. HBZ denotes half of the Brillouin zone, and the line integral is along the boundary of the HBZ. In this approach, the following constraint needs to be imposed on the wave function for the line integral: $|u_n(-\bm{k})\rangle=\Theta |u_n(\bm{k})\rangle$, where $\Theta$ is the time reversal operator. The $\mathbb{Z}_2=1$ state is topologically distinct from the $\mathbb{Z}_2=0$ state, in that they cannot be adiabatically connected to each other (through tuning some system parameter) without closing the bulk gap. The $\mathbb{Z}_2=0$ state is connected to the trivial vacuum, hence is the trivial insulating state. The $\mathbb{Z}_2=1$ state is nontrivial, and is known as the QSH state. It has an odd number of Kramers pairs of helical edge channels that are protected by time reversal symmetry.

\begin{figure*}
  \includegraphics[width=16cm]{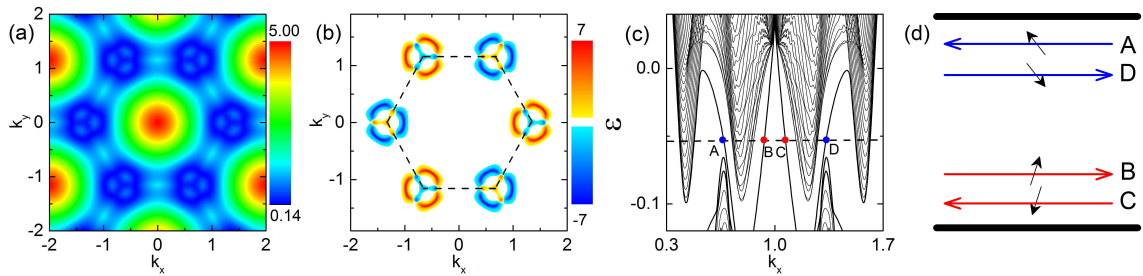}
  \caption{\label{FIG:Z2Metal} (color online) (a) Map of direct energy gap in $k$-space, (b) Berry curvature distribution in the Brillouin zone, and (c) the spectrum of a zigzag-terminated ribbon for a typical the $\mathbb{Z}_2$-metallic state corresponding to the marked point in the phase diagram Fig.~\ref{FIG:PhaseISB} with $t_R=0.3$ and $t_I=0.06$. The colored points in (c) label the edge modes localized at opposite boundaries, as schematically shown in
  (d) along with their spin polarizations. }
\end{figure*}

It has been shown that the single layer Kane-Mele model with only intrinsic SOC is a $\mathbb{Z}_2=1$ QSH ground state. This phase persists with finite Rashba SOC as long as the bulk gap does not close. When further increasing Rashba SOC, the gap finally closes and the QSH phase is destroyed. For a bilayer Kane-Mele model with each layer being a $\mathbb{Z}_2=1$ insulator, the combined system is in fact trivial because $1+1=0$ (mod 2). The case of bilayer model with only Rashba SOC has been studied before. It has been found that under finite interlayer potential $U$, the Rashba SOC in fact helps to realize a QSH state.

When the intrinsic SOC comes into the play, from previous discussion one expects that it tends to drive the system into the trivial phase, hence it has a competition with the Rashba SOC in the topological phase transition. Here we calculate the topological phase diagram reflecting the competition between the two SOCs. The result is shown in Fig.~\ref{FIG:PhaseISB} as functions of $t_\text{R}$ and $t_\text{I}$. The boundary between different topological phases corresponds to the states where the conduction band and the valence band touch. The black lines in the diagram mark such band-touching states. Here a finite interlayer potential $U$ is taken to open a bulk gap at first place. We observe that the phase diagram is divided by the band-touching lines into three regions. As expected, for large $t_\text{I}$, when intrinsic SOC dominates over Rashba SOC, the system takes a $\mathbb{Z}_2=0$ trivial phase (the upper part labeled as region III). For the opposite case, when Rashba SOC dominates over intrinsic SOC for large $t_\text{R}$, the system takes a $\mathbb{Z}_2=1$ QSH phase (the lower right part labeled as region II). One notes that the band-touching lines separates out another region at the lower left corner (region I) when both $t_\text{I}$ and $t_\text{R}$ are small, which also has $\mathbb{Z}_2=0$. Although both I and III are trivial in terms of $\mathbb{Z}_2$ classification, region I differs from III in the valley Chern number. Strictly speaking, the Chern invariant (number) is defined only for a closed manifold,\cite{Thouless,NiuQ}
\begin{equation}\label{EQ:CN}
\mathcal{C}=\frac{1}{2\pi}\sum_n\int d^2k\,\mathit \Omega_z(\bm k),
\end{equation}
where the integral is usually over the Brillouin zone. Because Berry curvature is odd under time reversal, the total Chern number must vanish in our present case. For a honeycomb lattice, the Brillouin zone has a hexagon shape and the energy spectrum has two valleys K and K' at the corners of the Brillouin zone, where the Berry curvature $\mathit \Omega_z$ is concentrated. In Fig.~\ref{FIG:Z2Metal}(b), we plot a typical Berry curvature distribution in $k$-space. One observes that the curvature is peaked around the valleys, and it has different sign between K and K' valleys. Therefore one can define a valley Chern number $\mathcal{C}_v$ as the difference between the integrals of the Berry curvature around the two valleys. For the three regions in Fig.~\ref{FIG:PhaseISB}, we have $\mathcal{C}_v=4$ for region I, $\mathcal{C}_v=2$ for region II, and $\mathcal{C}_v=0$ for region III. The valley Chern number physically determines the quantum valley Hall (QVH) conductance of the system.\cite{XiaoD1,YaoW1,YaoW2} Hence region I corresponds to a QVH phase while region III is valley Hall trivial.

What is more interesting of the present system is that we find there are extended regions around the phase boundaries in which the states are in fact metallic. The global bulk gap closes in the colored regions in Fig.~\ref{FIG:PhaseISB}. Because a metal does not have a stable ground state against excitations, usually it does not permit a topological phase. At the first sight, one might guess that the colored area in region II is just a trivial metallic phase, for which topological invariant cannot be defined. However, we find that although these states do not have a global bulk gap, they do have a local bulk gap at every $k$-point in the Brillouin zone. In Fig.~\ref{FIG:Z2Metal}(a), we plot the map of local direct gap of such a metallic state. It shows that the local gap is always bigger than zero. This means that the valence bands are still well-separated from the conduction bands and form an isolated manifold. They are adiabatically connected to the insulating state in each region without closing the local bulk gap (band-touching), hence also share the corresponding topological invariants.

We are most interested in the metallic states in region II. From the above argument, the metallic states there should have $\mathbb{Z}_2=1$ with spin helical edge channels, representing a 2D $\mathbb{Z}_2$-metallic phase. In Fig.~\ref{FIG:Z2Metal}(c), we show the energy spectrum calculated for a ribbon geometry of the system, corresponding to the marked point in the phase diagram. We observe that the band gap closes indirectly and there are edge states connecting between the conduction band and the valence band. We analyze their spatial distributions and spin polarizations. The result is plotted schematically in Fig.~\ref{FIG:Z2Metal}(d). From this figure, we can clearly observe that on each edge, there is one Kramers pair of spin helical edge states, as dictated by the $\mathbb{Z}_2=1$ requirement. As long as the time reversal symmetry is preserved, the two states on the same edge cannot be mixed with each other. However, scattering with finite momentum transfer may couple these state with the bulk states.

Such $\mathbb{Z}_2$-metallic state is analogous to the 3D Sb crystal. It has been found that the band gap of Sb also closes indirectly and its ground state has a nontrivial 3D $\mathbb{Z}_2$ invariant\cite{fu2007}. There are two differences. One is that the current $\mathbb{Z}_2$-metallic state is realized in a 2D system. The other one is its emergence here is a result of the competition between two types of SOC. As observed from the phase diagram, it requires both $t_\text{I}$ and $t_\text{R}$ to have finite values. The unique feature of this $\mathbb{Z}_2$-metallic state is the existence of topologically protected spin helical edge channels along with a metallic bulk.

Similar arguments apply to the metallic states in region I and region III as well. Those metallic states also maintain a local gap across the Brillouin zone. They are adiabatically connected to the corresponding insulating states in each region, hence share the same topological properties of those insulating states. Therefore, the colored area in region III is a trivial metallic phase, while the colored region in region I is a QVH metallic phase.

\section{Chern metallic phase}
\begin{figure}
  \includegraphics[width=8cm]{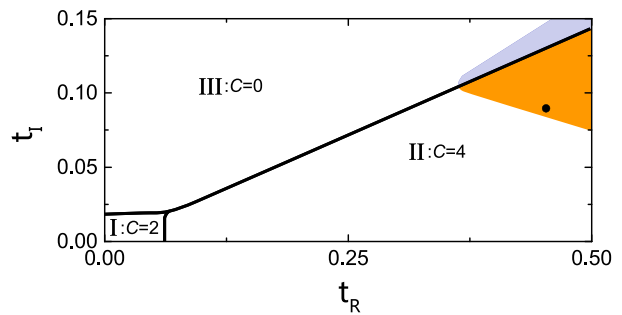}
  \caption{\label{FIG:PhaseTSB} (color online) Phase diagram for the bilayer Kane-Mele model with time reversal symmetry breaking. The solid lines separate three phases with different Chern numbers $\mathcal{C}$. The dot marks a Chern metallic state. The parameters used for the calculations are $t=1$, $M=0.1$, $U=0$, and $t_{\perp}=0.2$. }
\end{figure}
\begin{figure*}
  \includegraphics[width=16cm]{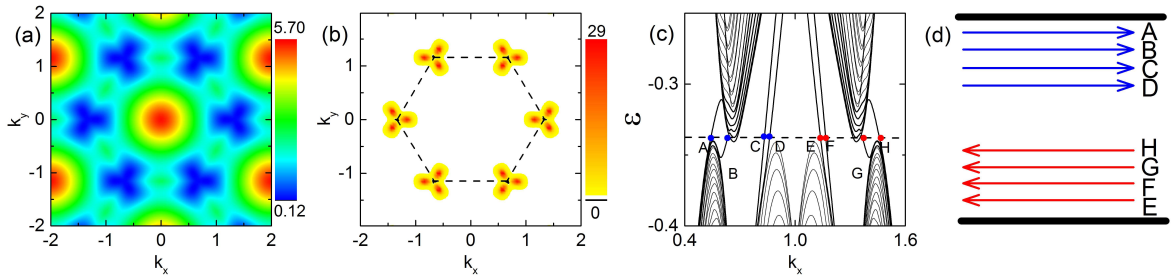}
  \caption{\label{FIG:ChernMetal} (color online) (a) Direct energy gap, (b) Berry curvature distribution in the Brillouin zone, and (c) the spectrum of a zigzag-terminated ribbon for a typical Chern metallic state as marked in the phase diagram Fig.~\ref{FIG:PhaseTSB} with $t_R=0.45$ and $t_I=0.09$. The colored points in (c) label the edge modes localized at opposite boundaries, as shown schematically in
  (d). }
\end{figure*}

When time reversal symmetry is broken, the topological classification of a 2D insulator changes from $\mathbb{Z}_2$ to $\mathbb{Z}$, characterized by the Chern number. The Chern number, given by the integral of Berry curvature over the entire Brillouin zone, determines the number of chiral edge channels of the system.  For our bilayer Kane-Mele model, the time reversal symmetry is broken by a finite $M$ value, which generates a Zeeman spin-splitting. Here we set $U=0$ such that the system preserves inversion symmetry. Following the same methodology as in the previous section, we calculate the phase diagram with respect to the parameters $t_\text{R}$ and $t_\text{I}$. In Fig.~\ref{FIG:PhaseTSB}, the phase boundaries, which are the band-touching lines, separate the phase diagram into three regions. The region I at the lower left corner with both $t_\text{I}$ and $t_\text{R}$ small is an insulating phase with Chern number $\mathcal{C}=2$. In region II, $t_\text{R}$ dominates over $t_\text{I}$, and we have $\mathcal{C}=4$. This is consistent with the previous observation that $\mathcal{C}=2$ for a single layer with only Rashba SOC. In region III, $t_\text{I}$ dominates over $t_\text{R}$, and we have $\mathcal{C}=0$ indicating that it is a trivial phase.

If we analyze the spectrum of the system, we find that there are also extended areas in region II and region III for which the global gap closes but local direct gap is still maintained. This is demonstrated in Fig.~\ref{FIG:ChernMetal}(a) for a typical case, in which one observes that the direct gap is always positive throughout the Brillouin zone. Due to the presence of the local gap, the valence bands are separated from the conduction bands. These metallic states are adiabatically connected to the insulating states in the same region. The metallic states in region III are trivial, while the metallic states in region II are topologically nontrivial.

Let's focus on the states in region II, which has a Chern number $\mathcal{C}=4$. The adiabatic connectivity means that the metallic state shares the same topological character as the Chern insulator state: it also has chiral edge states and a Chern number can be defined for its valence bands with $\mathcal{C}=4$. We shall call these novel states as Chern metals.

In Fig.~\ref{FIG:ChernMetal}(b), we show the Berry curvature distribution in the $k$-space for the $\mathcal{C}$ metallic state marked by the dot in the phase diagram. It is observed that the Berry curvature is concentrated in each valley, and it has the same sign for different valleys, obeying the requirement from inversion symmetry. In Fig.~\ref{FIG:ChernMetal}(c), we plot the energy spectrum of a ribbon geometry for the Chern metallic state. We observe that the global bulk gap closes indirectly but the chiral edge states still exist. We analyze the spatial distribution of these edge states and the result is shown schematically in Fig.~\ref{FIG:ChernMetal}(d). On each edge, there are four channels propagating along the same direction, as required by the Chern number $\mathcal{C}=4$. Because it is a metallic state, the chiral edge states are not so robust as those in an insulating state because scattering may couple these edge states with the bulk states. The existence of chiral edge channels along with a conducting bulk is the hallmark of this novel Chern metallic phase. From the phase diagram Fig.~\ref{FIG:PhaseTSB}, we see that its appearance must require both $t_\text{R}$ and $t_\text{I}$ to have a finite value. Hence like the $\mathbb{Z}_2$-metallic state, it is a result of the competition between intrinsic SOC and Rashba SOC.

\section{Summary}

In this work, we have studied the topological phases of a bilayer Kane-Mele model in detail. We find that the system exhibits a rich phase diagram as a result of the competition between intrinsic SOC and Rashba SOC. In contrast to the single layer case, there emerge novel 2D topological metallic phases that can be adiabatically connected to their topological insulator counterparts. In the presence of time reversal symmetry, we find there is $\mathbb{Z}_2$-metallic phase with nontrivial $\mathbb{Z}_2$ invariant and spin helical edge states. When time reversal symmetry is broken, there exists Chern metallic phase with nontrivial Chern invariant and chiral edge states. These findings broaden our knowledge of the topological states of matter. It also shows that by controlling the competition between different SOCs, one can tune the topological phase and control both the bulk and edge conduction, which could be useful in applications based on topological materials.

\emph{Acknowledgement.}  This work was supported by SUTD-SRGEPD2013062, the MOST Project of China (Grants Nos. 2014CB920903 and 2011CBA00100), the NSFC (Grants Nos. 11174022, 61227902, 11174337, and 11225418), the NCET program of MOE (Grant No. NCET-11-0774), and the Specialized Research Fund for the Doctoral Program of Higher Education of China (Grant No. 20121101110046). Z.Q. was financially supported by USTC Startup, Bairen Program of CAS, and NNSFC (No. 91021019).

\end{document}